\documentstyle[onecolumn, psfig]{mn-1_4}
\def\beqa{\begin{eqnarray}}
\def\eeqa{\end{eqnarray}}
\let\gam=\gamma

\def\hm {$h^{-1}$ Mpc }

\def\cm{\rm cm}
\def\cm3{\rm cm^{-3}}

\def\spose#1{\hbox to 0pt{#1\hss}}
\def\approxlt{\mathrel{\spose{\lower 3pt\hbox{$\sim$}}\raise 2.0pt\hbox{$<$}}}
\def\approxgt{\mathrel{\spose{\lower 3pt\hbox{$\sim$}}\raise 2.0pt\hbox{$>$}}}

\begin{document}

\title{Weak Gravitational Lensing by Voids}
\author[L. Amendola, J. A. Frieman and I. Waga]{Luca Amendola$^{1, 2}$, 
Joshua A. Frieman$^{1,3}$ and Ioav Waga$^{1, 4}$ \\
$^1$NASA/Fermilab Astrophysics Center, Fermi National Accelerator Laboratory 
\\
PO Box 500, Batavia IL 60510 - USA\\
$^2$Osservatorio Astronomico di Roma\\
V. del Parco Mellini, 84, 00136 Rome - Italy\\
$^3$Department of Astronomy and Astrophysics, University of Chicago\\
5640 S. Ellis Avenue, Chicago, IL 60637 - USA\\
$^4$Universidade Federal do Rio de Janeiro, Instituto de F\'\i sica\\
Rio de Janeiro, RJ, 21945-970 - Brazil.}

\maketitle

\begin{abstract}
We consider the prospects for detecting weak 
gravitational lensing by underdensities
(voids) in the large-scale 
matter distribution. We derive the basic expressions for
magnification and distortion by spherical voids.  
Clustering of the background sources 
and cosmic variance are the main factors which limit in principle 
the detection of lensing by voids. We
conclude that only voids with radii larger than $\sim 100$ \hm  
have lensing signal to noise larger than unity.
\end{abstract}

\begin{keywords}
gravitational lensing: voids : weak lensing
\end{keywords} 

\section{Introduction}

Gravitational lensing remains a rapidly evolving field of study, 
with observational evidence now firmly established for the phenomena of 
strong, weak, and micro-lensing. 
Examples include multiple imaging of background quasars by 
foreground galaxies, weak distortion of background galaxy images by 
foreground galaxy clusters, and time-dependent enhancement of stellar 
brightness by underluminous objects in our own galaxy.

In each of these cases, the 
intervening matter between the source and observer acts as a converging
lens, leading to image splitting, magnification, and distortion. 
In addition to mass concentrations,  
{\it underdensities} in the matter distribution---voids---also affect 
the propagation of light rays across cosmological distances. 
Over the last fifteen years, galaxy redshift 
surveys have revealed that a substantial fraction of the 
volume of the Universe is occupied by giant
voids, quasi-spherical regions almost empty of luminous matter
(e.g., de Lapparent {\it et al.} 1989, El-Ad {\it et al.} 1996, Da
Costa {\it et al.} 1996). However, relatively little attention has been 
devoted to the possibility of detecting the gravitational lensing 
effects of voids. A light ray passing through an empty region 
is deflected away from the center of the void, since it is 
gravitationally attracted by the surrounding matter. This
phenomenon is mentioned, although not discussed, in Hammer \& Nottale
(1986), in connection with a cluster model with a surrounding underdense
shell. It is also discussed in the linear limit by Moreno \& Portilla
(1990), but they do not draw observational conclusions. The 
effects of large empty regions on light propagation in 
FRW universes has been extensively studied (e.g., Dyer \& Roeder 1972; 
see Holz \& Wald 1998 for a recent discussion), but these 
works did not focus on the detectability of individual voids 
via lensing.

In this paper we study gravitational lensing by large-scale voids and the possibility of 
its detection. Our motivations are several: First,
voids are ubiquitous and large, accounting for some 80\% of the volume of the
Universe. It is in fact rather unlikely for a light ray from a distant source 
{\it not} to have passed through a void. Second, to
understand the formation of such large voids we would like to know their
matter content: are they nearly devoid of matter or only lacking bright
galaxies? In principle, gravitational lensing can probe the
matter content of voids. Third, some theories predict the formation of {\it 
primordial voids}, e.g., due to cosmic explosions (Ostriker \& Cowie 1981;
Yoshioka \& Ikeuchi 1989) or primordial phase transitions (La 1991;
Amendola \& Occhionero 1993; Amendola {\it et al.} 1996). A search for such
primordial structures could be pursued via  their gravitational
lensing signature. Fourth, ``negative" masses, i.e., underdense regions, lead 
to new lensing phenomena that are of intrinsic theoretical interest.

Thus, we pose the following question: suppose a void of scale $R$ at redshift 
$z_L$ is identified in a galaxy redshift survey; can its lensing effect 
on background sources be measured and use to probe its matter density?
Since this is an exploratory study, we consider a highly idealized model 
of a void, comprising a uniform spherical underdensity surrounded by 
a uniform overdense shell, in an otherwise homogenous universe. While only 
a starting point for a more realistic treatment, such a model captures 
the broad features found in numerical and analytical studies of voids 
growing via gravitational instability. We consider two techniques for 
detecting the lensing effect of voids, color-dependent
angular density amplification (Broadhurst, Taylor \& Peacock 1995, 
Broadhurst 1996) and aperture densitometry (Kaiser 1994,
Schneider 1996). Unfortunately, we find that only voids with radii 
larger than 100 \hm can be individually detected via weak lensing; 
empty voids with radii $R=30$ \hm, characteristic of those seen in 
galaxy redshift surveys, have a lensing signal-to-noise ratio
smaller than unity. 

This paper is organized as follows: 
In Sec. $2$ we review the basic lens theory and apply it to lensing by spherical voids. The possibility of detecting the lensing effect using the color-dependent angular density amplification is investigated in Sec. $3$. In Sec. $4$ we apply the technique of aperture densitometry  to the detection of voids. Our main conclusions are stressed out in Sec. $5$.

\section{Lens theory and the spherical void}

Here we review the basic features of gravitational lens theory and 
apply it to lensing by spherical voids.
The lensing geometry in the case of voids is shown in Figure 1 . In
the figure, L is the center of the void, S is a distant point source, and O the
observer. The source is at distance $D_{OS}$ from the observer and $D_{LS}$
from the lens, and the distance between lens and observer is denoted by 
$D_{OL}$; these are angular diameter distances.
Light rays from the source are `repelled' by 
the lens center and are deflected by the angle $\hat{\alpha}$ (the deflection angle
is negative for a matter underdensity). The angular
separations between lens and image and between lens and source are denoted by $\hat{
\theta}_I $ and $\hat{\theta}_S $. We note that a void can
be thought of as an ``extended negative mass lens'' and some of its lensing
properties can be obtained from that of a ``negative point mass lens'' in exactly the
same way as the standard point mass lens is generalized to an extended lens
(see, e.g., Schneider, Ehlers and Falco (1992)). Throughout this paper we
adopt the ``thin lens'' approximation, applicable if the void radius $R$
is small compared with $D_{OL}$ and $D_{LS}$.

Now consider photons traveling in a weak gravitational field generated by a
stationary bound system. If the deflection angle is small, it can be expressed
as 
\begin{equation}
\hat{\alpha} = \frac{2}{c^2} \int_{-\infty}^{\infty}\;\vec{\nabla}
\Phi_{_{3D}}\; dl,
\end{equation}
where $\Phi_{_{3D}}$ is the 3-dimensional Newtonian potential, satisfying
the boundary conditions $\Phi =0$ and $\vec{\nabla} \Phi_{_{3D}} =0$ at
infinity. Eqn. (1) follows from linearized general relativity; to lowest order, 
the integral is performed along the unperturbed (zero field) photon trajectory. 
In this approximation, $\hat{\alpha}$ is contained in the lens plane.

\begin{figure}
\hspace*{0.15in} \vspace*{0.5cm} 
\psfig{file=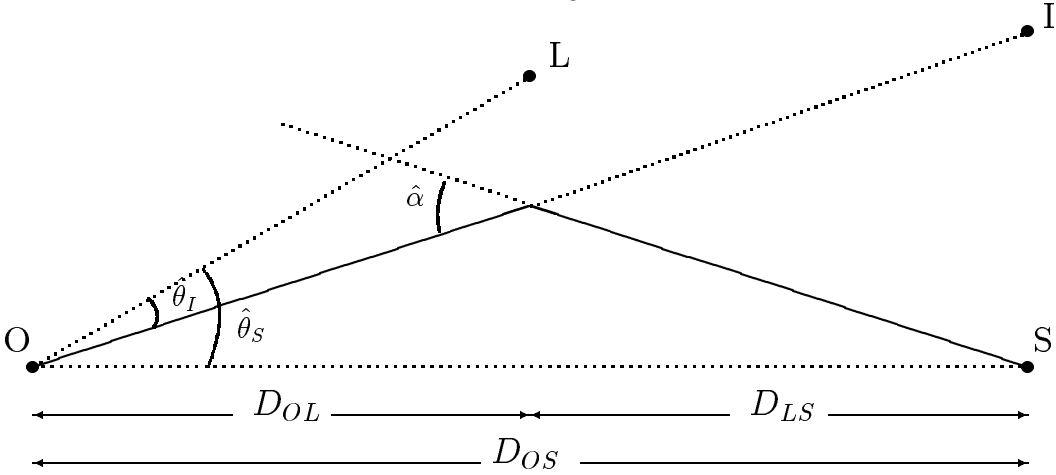,height=5.8cm,width=14cm}
\caption{Void lensing geometry. Light rays from the source S are deflected 
by the lens L through angle $\hat{\alpha}$ and detected by observer O as an 
image at position I.}
\end{figure}

From Figure 1 we obtain the lens equation 
\begin{equation}
\hat{\theta} _S=\hat{\theta} _I-\frac{D_{LS}}{D_{OS}}\hat{\alpha}~.
\end{equation}
Introducing rescaled two-dimensional coordinates in the lens plane, 
$x_j=D_{OL}{\theta _I}_j$ and $
y_j=D_{OL}{\theta _S}_j$,  from Eqn.(2) we obtain 
\begin{equation}
\frac{\partial y_i}{\partial x_j}=\delta _{ij}-\frac{D_{OL}D_{LS}}{D_{OS}} 
\frac{\partial \alpha _i}{\partial x_j}.
\end{equation}
Using Eqn.(1), we can rewrite Eqn.(3) as 
\begin{equation}
\frac{\partial y_i}{\partial x_j}=\delta _{ij}-\Psi _{ij},
\end{equation}
where the 2-dimensional potential $\Psi $ is given by 
\begin{equation}
\Psi =\frac 2{c^2}\frac{D_{OL}D_{LS}}{D_{OS}}\int_{-\infty }^\infty \Phi
_{_{3D}}dx_3
\end{equation}
and $\Psi _{ij}\equiv \partial^2\Psi/\partial x_i\partial x_j$.

Since $\Phi _{_{3D}}$, $\vec{\nabla}\Phi _{_{3D}}$ $
\rightarrow 0$ at infinity, from Poisson's equation we find  
\begin{equation}
\int_{-\infty }^\infty \left(\frac{\partial ^2\Phi _{_{3D}}}{\partial {x_1}^2}
+ \frac{\partial ^2\Phi _{_{3D}}}{\partial {x_2}^2}\right)
dx_3\equiv 4\pi G\;\frac{\Sigma _c}2\nabla _{_{2D}}^2\Psi =4\pi G\;\Sigma ,
\end{equation}
where $\Sigma \equiv \int_{-\infty }^\infty \;\rho \;dx_3$ is the surface
mass density, and the critical surface density $\Sigma _c$ is defined by 
\begin{equation}
\Sigma _c \equiv \frac{c^2}{4\pi G}\frac{D_{OS}}{D_{OL}D_{LS}}.
\end{equation}
Defining the convergence, $\kappa \equiv \Sigma/ \Sigma_c$, it
follows from Eqn.(6) that $\nabla _{_{2D}}^2\Psi =2\kappa .$

The shear is defined by 
 
\begin{equation}
\gamma^2 = {\gamma _1}^2+{\gamma _2}^2,
\end{equation}
where $\gamma _1=\frac 12(\Psi _{11}-\Psi _{22})$ and $\gamma _2=\Psi
_{12}=\Psi _{21}.$ From Eqn.(4), 
the Jacobian matrix of the lensing mapping, which describes its local 
properties, can be expressed as  
\begin{equation}
A^{-1}=\pmatrix{1-\kappa - \gamma_1 & -\gamma_2&\cr \noalign{\smallskip}
-\gamma_2 & 1-\kappa + \gamma_1\cr}~.
\end{equation}
Since lensing conserves surface brightness, 
the image magnification $\mu $ is given by the increase in image area 
due to convergence and shear, 
\begin{equation}
\mu =\frac 1{\det A^{-1}}=\frac 1{(1-\kappa )^2-\gamma ^2}.
\end{equation}

We now apply this formalism to the spherical void lens. We model 
the void as a spherically symmetric matter distribution, with 
density excess $\delta \rho_v = \rho_v - \bar \rho$ given by  
\begin{eqnarray}
\delta \rho_v (r) = \left\{ 
\begin{array}{lll}
- \bar \rho \delta & \;\;{\rm for}\;\;\;\;\;\;\;\;\;\;\; r <R, &  \\ 
\rho_{+} \delta & \;\;{\rm for}\;\;\;\; R<r<R+d, &  \\ 
0 & \;\;{\rm for}\;\; \;\;\;\;\;\;\;\;\;\; r>R+d. & 
\end{array}
\right.
\end{eqnarray}
Here, $\delta \rho_v/\bar \rho$ is the density contrast ($\delta < 1$), and 
$\bar \rho $ is 
the mean energy density of the Universe. In Eqn.(11), $R$ is the
void radius, and $d$ is the thickness of the surrounding mass shell. We assume that the
mass excess in the shell exactly compensates the amount missing in the void, 
so $\rho _{+}$ and $\bar \rho $ are related by  
\begin{equation}
\rho _{+}=\frac{\bar \rho }{\left(1+\frac dR \right)^3-1} ~~.
\end{equation}
In the limit $d\rightarrow \infty $ we obtain $\rho_{+}\rightarrow
0 $, and we recover the simple uncompensated void.

It is convenient to define the impact parameter $b$ as the distance of 
closest approach of the unperturbed ray to the lens L.
Introducing the scaled distances $\tilde b=b/R$ and $\tilde d=d/R$, the 
the surface mass density for the void model of Eqn.($11$) takes the form, 
\begin{eqnarray}
\Sigma (\tilde b) = \left\{ 
\begin{array}{lll}
- 2 \bar \rho \delta R \; \left[ \left( 1-\tilde b^2 \right)^{1/2}+ \frac{
\left( 1-\tilde b^2 \right)^{1/2}-\left((1+ \tilde d)^2 - \tilde
b^2\right)^{1/2}}{(1+\tilde d)^3 -1}\right] & \;\;\;{\rm for}
\;\;\;\;\;\;\;\;\;\;\;\;\;\tilde b <1, &  \\ 
2 \bar \rho \delta R\;\frac{[ (1+\tilde d )^2-\tilde b^2 ]^{1/2}} {(1+\tilde
d)^3 -1} & \;\;\;{\rm for}\;\;\;\;\;\; 1<\tilde b<1+\tilde d, &  \\ 
0 & \;\;\;{\rm for}\;\; \;\;\;\;\;\;\;\;\;\;\;\tilde b>1+\tilde d. & 
\end{array}
\right.
\end{eqnarray}

The projected effective mass interior to the impact parameter $b$, ${\it M}
(b)=2\pi \int_0^ba\Sigma (a)da$, is given by, 
\begin{eqnarray}
{\it {M}(\tilde{b})=\left\{ 
\begin{array}{lll}
-\frac{M_v}{\tilde{d}^3+3\tilde{d}^2+3\tilde{d}}\left\{ \left[ (1+\tilde{d}
)^2-\tilde{b}^2\right] ^{3/2}-(1+\tilde{d})^3(1-\tilde{b}^2)^{3/2}\right\} & 
\;\;\;{\rm for}\;\;\;\;\;\;\;\;\;\;\;\tilde{b}<1, &  \\ 
-\frac{M_v}{\tilde{d}^3+3\tilde{d}^2+3\tilde{d}}\left[ (1+\tilde{d})^2-
\tilde{b}^2\right] ^{3/2} & \;\;\;{\rm for}\;\;\;\;1<\tilde{b}<1+\tilde{d},
&  \\ 
0 & \;\;\;{\rm for}\;\;\;\;\;\;\;\;\;\;\;\tilde{b}>1+\tilde{d}, & 
\end{array}
\right. }
\end{eqnarray}
where $M_v=\frac 43\pi \bar{\rho}\delta R^3$.

The magnification factor for a spherical void takes a simple form, 
\begin{equation}
\mu = \left[1-\frac{\Sigma_v}{\Sigma_c} \frac{\tilde M(\tilde b)}{{\tilde b}
^2}\right]^{-1} \left[1- \frac{\Sigma_v}{\Sigma_c}\frac{d}{d\tilde b}\left ( 
\frac{\tilde M(\tilde b)}{\tilde b}\right) \right]^{-1},
\end{equation}
where $\tilde M(\tilde b)=M(\tilde b)/M_v$ and $\Sigma_v = M_v/\pi R^2$.

In what follows we shall assume the Einstein-de Sitter universe ($\Omega_m =1 $), 
in which the angular diameter distance takes a simple form,  
\begin{equation}
D_{LS}=r(z_L,z_S) = c{H_0}^{-1}\frac 2{1+z_S}\times
\left[(1+z_L)^{-1/2}-(1+z_S)^{-1/2}\right] ,
\end{equation}
and $D_{OL}=r(0,z_L)$, $D_{OS}=r(0,z_S)$.

The dimensionless quantity, $\Sigma_v/\Sigma_c$, which depends on $z_S$, $z_L$, $\delta$, 
and $R$, clearly sets the strength of gravitational lensing by voids. For
instance, the magnification $\mu $ can be written in the limit of weak
lensing as $\mu =1+(\Sigma_v/\Sigma_c)f(\tilde{b},\tilde{d})$, where 
$f(\tilde{b},\tilde{d})$ is a function of order unity. For the 
Einstein-de Sitter model, 
\begin{equation}
\frac{\Sigma_v}{\Sigma_c} = \frac{2 R \delta D_{OL}D_{LS}}{(c/H_0)^2 D_{OS}} ~,
\end{equation}
where $c/H_0 = 3000$ \hm is the Hubble radius. For voids at cosmological 
distances, $\Sigma_v/\Sigma_c \sim R\delta/cH_0^{-1}$; for example, 
for $R=50h^{-1}{\rm 
~Mpc}$, typically $\sum_v/\sum_c=.003-.005$. The ratio $\sum_v/\sum_c$ 
reaches at most 1\% for giant voids of radius 100 $h^{-1}{\rm ~Mpc}$.

\begin{figure}
\hspace*{0.15in}
\psfig{file=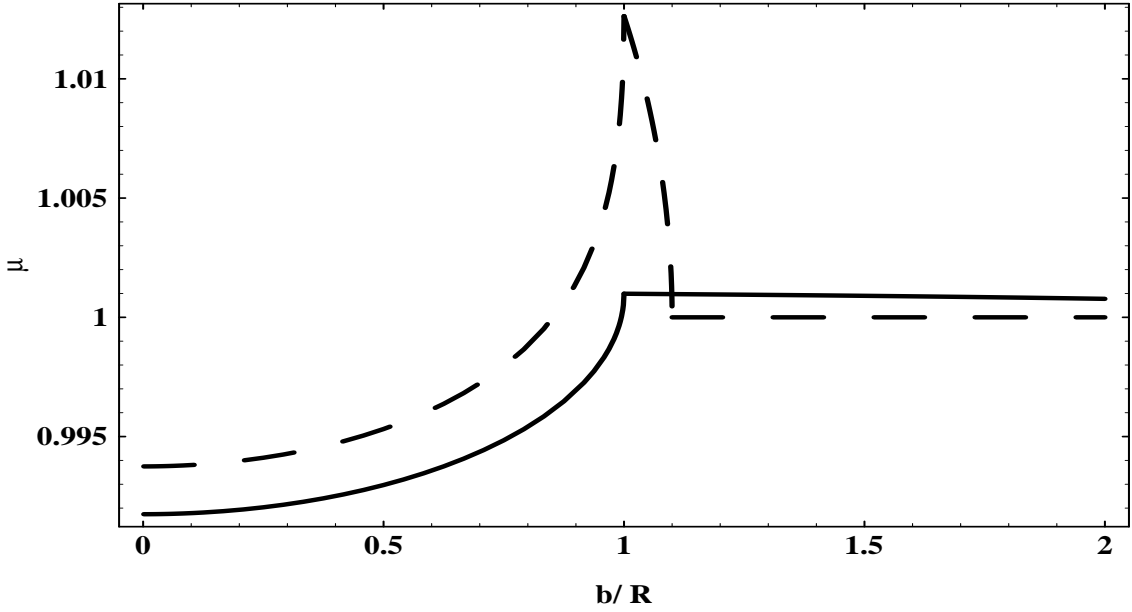,height=8cm,width=15cm}
\caption{The magnification $\mu$ versus scaled impact parameter 
$\tilde b$ in for two representative voids. The
dashed curve corresponds to a compensated void with $\Sigma_v/\Sigma_c 
= 0.005$ and $\tilde d =0.1$. The continuous curve is for an uncompensated
void with $\Sigma_v/\Sigma_c = 0.005 $ and $\tilde d =2$.}
\end{figure}

In Fig. 2 we show the magnification $\mu $ as a function of the 
scaled impact parameter $\tilde{b}$ for two different types of voids. The dashed
curve corresponds to a compensated void with $\Sigma_v/\Sigma_c=0.005$ and $\tilde{
d}=0.1$. We also show the almost uncompensated void with $
\Sigma_v/\Sigma_c=0.005 $ and $\tilde{d}=2$. In the first case there is a peak
at the void boundary ($\tilde{b}=1$) where images are always magnified ($\mu >1$). 
Images formed well inside the void ($\tilde{b} \ll 1$) are demagnified ($\mu <1$);  
for $\tilde{b}\geq 1+\tilde{d}$ there is no magnification. From Fig.1, 
the demagnification inside the void is higher in
the case of an uncompensated void. In the weak lensing limit, $\kappa, \gamma \ll 1$, 
the magnification
depends only on the convergence, $\mu \simeq 1+2\kappa$.

We now consider the image distortion induced by a void. A
small circular source will be compressed by a factor $
f_r=[1-(\kappa-\gamma)]^{-1}$ along the radial direction (away from the void
center) and by a factor $f_o=[1-(\kappa+\gamma)]^{-1}$ along the tangential 
direction. For a spherical void, the convergence and  shear can be
expressed as 
\begin{eqnarray}  \label{kg}
\kappa&=&\frac12\frac{\Sigma_v}{\Sigma_c} \left[\frac{\tilde M(\tilde b)}{
\tilde b ^2}+\frac{d}{d\tilde b}\left(\frac{\tilde M(\tilde b)}{\tilde b}
\right)\right] \,,  \nonumber \\
\gamma &=& \frac12\frac{\Sigma_v}{\Sigma_c} \left[\frac{\tilde M(\tilde b)}{
\tilde b ^2}-\frac{d}{d\tilde b}\left(\frac{\tilde M(\tilde b)}{\tilde b}
\right)\right]\,.
\end{eqnarray}

\begin{figure}
\hspace*{0.15in} 
\psfig{file=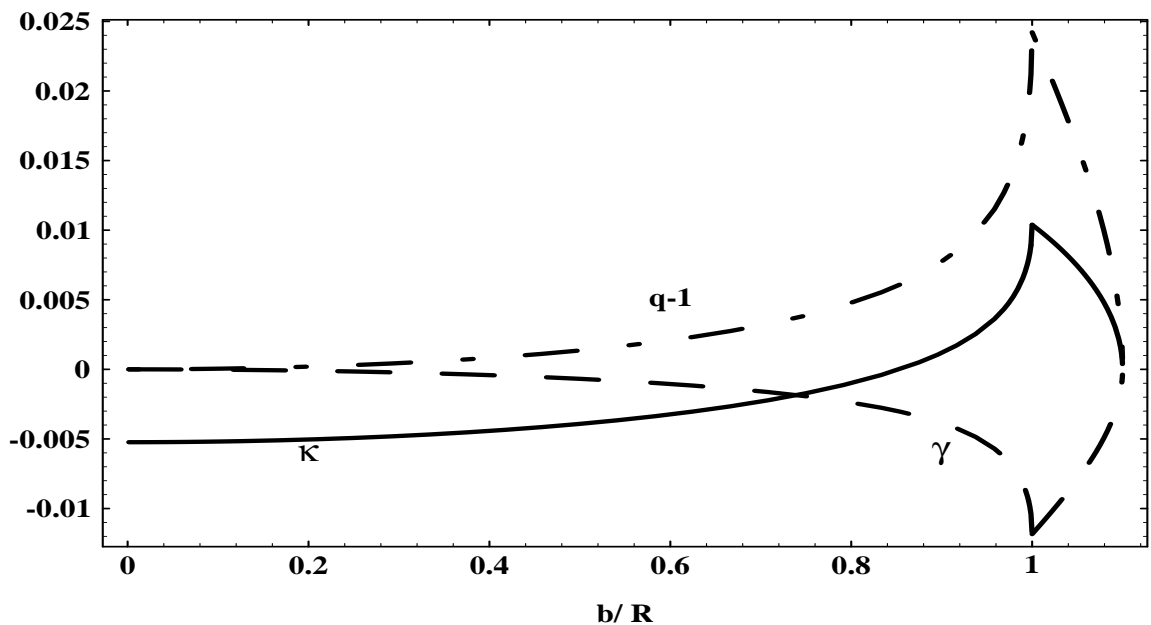,height=8cm,width=15cm}
\caption{The convergence $\kappa$, shear $\gam$, and ellipticity $q-1$ 
as a function of $\tilde b$, for $\Sigma_v/\Sigma_c = 0.005$ and $\tilde d =0.1$.}
\end{figure}

As a measure of the image distortion, 
it is useful to define the induced ellipticity $q$ for a circular source,  
\begin{equation}
q={\frac{1-(\kappa +\gamma )}{1-(\kappa -\gamma )}}\,.  \label{qdist}
\end{equation}
For small $\kappa $ and $\gam$, the ellipticity depends only on the 
shear, $q \simeq 1 - 2\gamma$. For negative shear, 
as is the case for voids, the radial image compression is smaller than the 
tangential compression; as a result, the images tend to be preferentially 
aligned in the radial direction. For positive mass concentrations (e.g., 
galaxy clusters), in the weak lensing limit the images are aligned tangentially.
Since our void model is spherical,
there is no rotation of the source image. In Fig. 3 we show $\kappa $, $\gam$,  
and $q-1$ as a function of $\tilde{b}$, for $\Sigma_v/\Sigma_c=0.005$ and $
\tilde{d}=0.1$. In Fig. 4, we display how randomly distributed circular background
galaxies would appear in the case of
void lensing (left panel) and in the case of an isothermal foreground cluster (right 
panel). In this figure, the source galaxies were all placed at the same 
redshift, and the distortion has been exaggerated to show the effect.

\begin{figure}
\hspace*{0.1in} 
\psfig{file=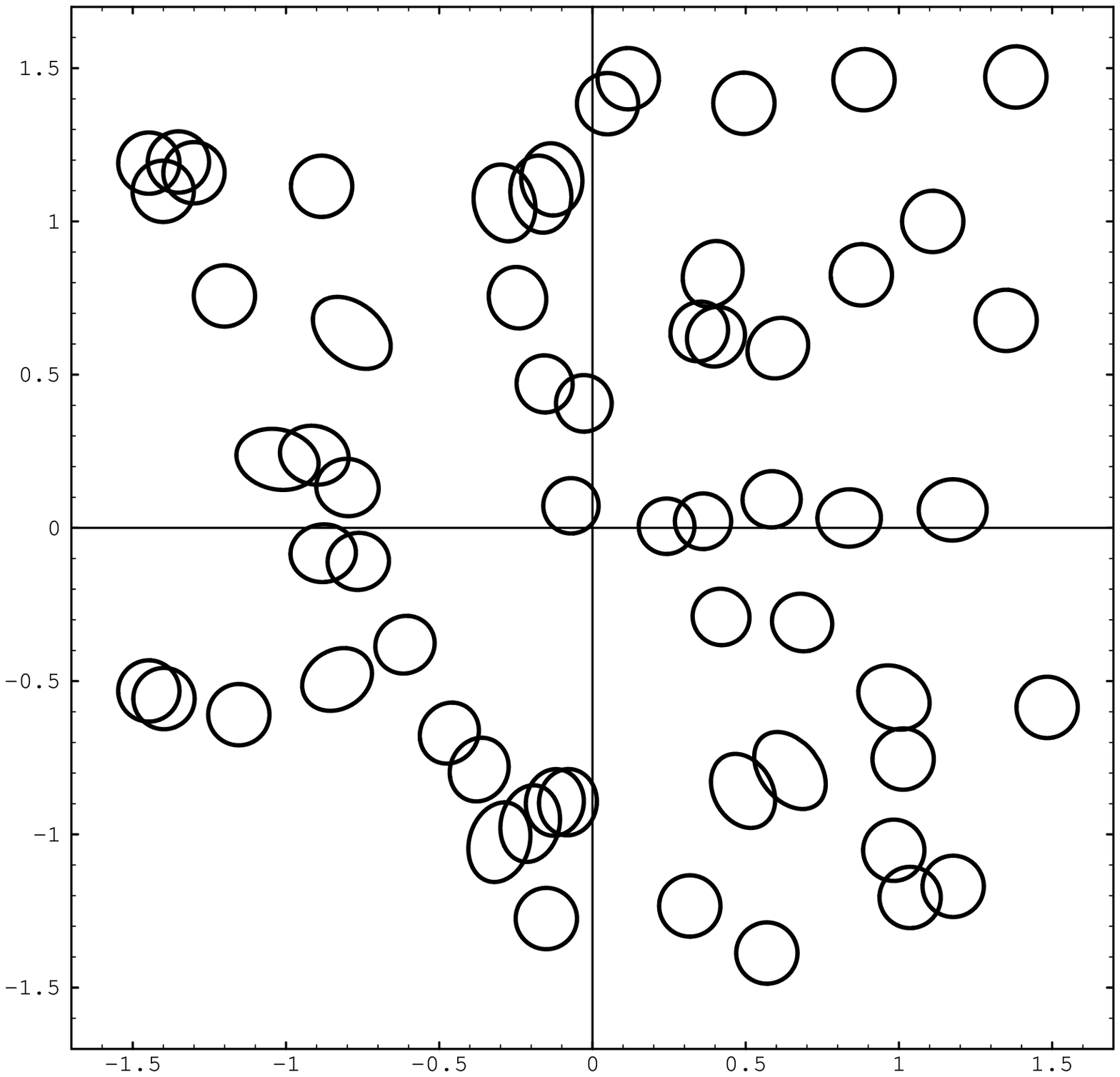,height=7.5cm,width=7.5cm}
\hspace*{0.2in} 
\psfig{file=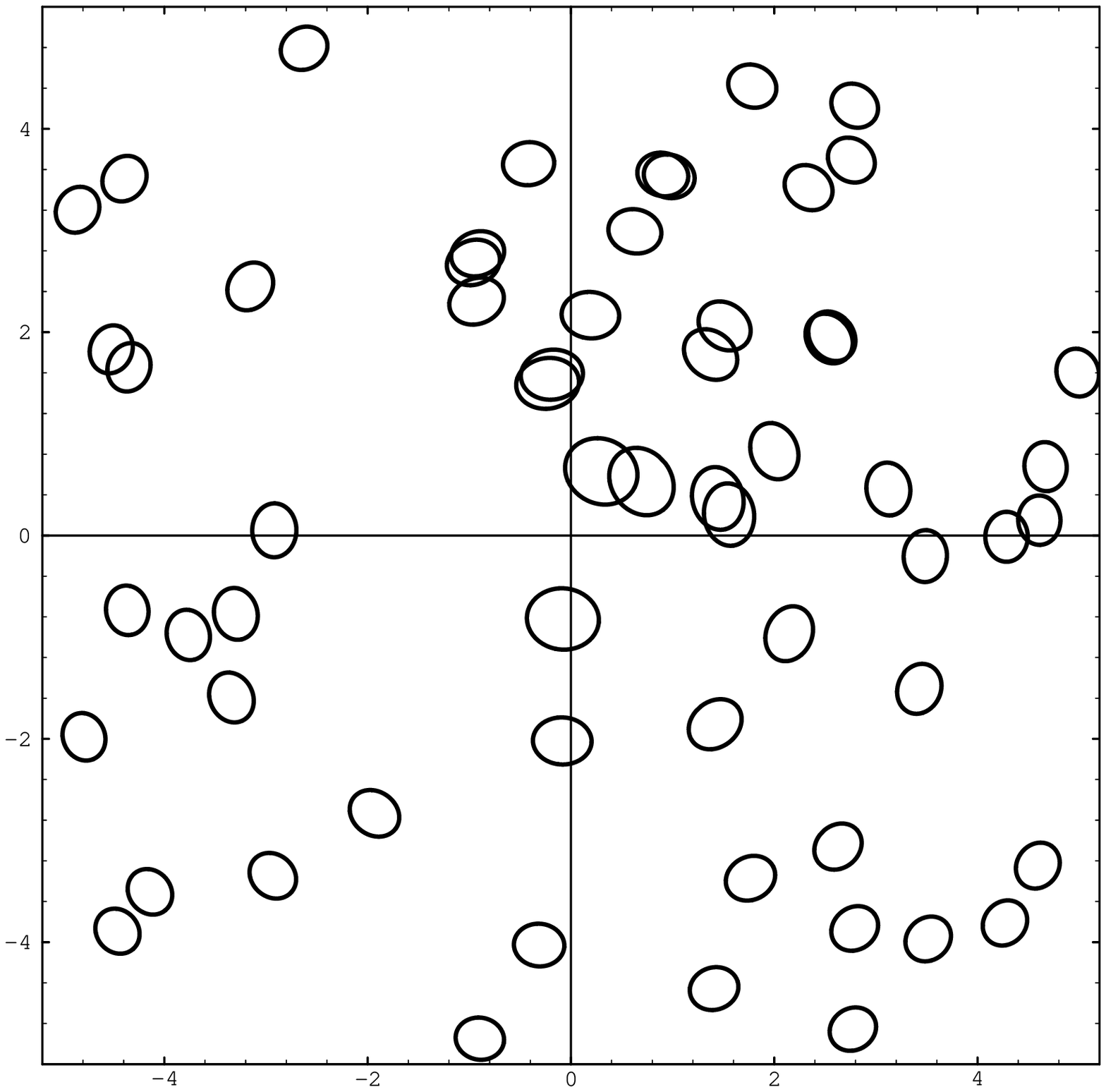,height=7.5cm,width=7.5cm}
\caption{The appearance (ellipticity) of randomly distributed circular background 
galaxies in the case of void lensing (left panel) and for an isothermal lensing cluster
(right panel). }
\end{figure}

\section{Color-dependent density magnification}
Since lensing changes the apparent brightness of a source, the 
galaxy number count vs. magnitude relation will be altered for 
sources behind a foreground lens; this effect has been proposed as a method 
for probing cluster masses (Broadhurst, Taylor \& Peacock 1994,  
Broadhurst 1996, Van Kampen 1998). Weak lensing causes two opposing
effects on the number density of background sources to fixed limiting magnitude.  
Compared to an unlensed sample, sources behind a void are 
demagnified by a factor $\mu <1$, resulting in fewer 
sources in a magnitude-limited survey. 
On the other hand, the radial deflection of light rays  
by the foreground void causes the background images to appear concentrated 
toward the line of sight to the lens, increasing their areal density by a factor $1/\mu$.
For galaxies observed in the red, the latter effect dominates, leading to an enhancement in 
the number density of galaxies behind a void; for lensing by clusters, one 
expects a suppression of red background source counts.

Let $n_0(z)$ and  $n_b(z)$ denote the number of galaxies per square degree per
redshift interval in an unlensed sample and in a lensed sample with 
characteristic magnification $\mu$ centered on a void.
Let $N_0(m)$ and $N_b(m)$ be the corresponding total number of galaxies per
square degree in each sample to a given limiting magnitude $m$.
Due to the two competing effects noted above, the counts are related by  
\begin{equation}
n_b/n_0=\mu ^{2.5s -1}~,
\end{equation}
where $s $ is the background galaxy count-magnitude slope, $s =d\log N_0/dm$.
Broadhurst (1994,1996) reports that $s \approx 0.5$ in $B$ 
and $s \approx 0.15$ in $R$ at magnitudes $m \leq 23$.
In the case of void lensing, except near the void boundary 
we have $\mu <1$, leading to an enhancement of galaxies observed in $R$ 
behind the lens than in the far field. 
In general, the slope $s$ varies with magnitude 
(Broadhurst, Taylor, \& Peacock 1995). To simplify the analysis, 
we will assume 
$s$ is constant and use the value $s= 0.15$ in numerical calculations. 
In fact this appears to be a good approximation up to $m_{lim} \sim 22$. For 
deeper surveys, the effective value of $s$ is larger and the effect we are 
estimating weaker. 
As will be clear from our results, even using this somewhat optimistic 
approximation,  
it is very difficult to obtain a high signal to noise ratio with this 
technique for 
voids.

Let us estimate the fractional variation $\delta _b=(N_b-N_0)/N_0$ in
the number count of red galaxies behind a void and compare it with various
sources of noise. We assume that the unlensed redshift distribution
of the sources in the sample is given by the normalized function
$p(z)$, which we will specify below.
In the weak lensing limit, the magnification $\mu =1+2\kappa$; 
therefore, for objects at fixed angular separation $\theta_S $
from the lens center at $z_L$  
\begin{equation}
\delta _b=\int_{z_L}^\infty (5s -2)\kappa(z_S,\theta_S) p(z_S) dz_S ~.
\end{equation}

There are four main sources of statistical error in this measurement: the variance $
\sigma _L$ due to lensing by random density fluctuations along the line of
sight, the Poisson noise $\sigma _P$ due to the finite number of sources, the variance $
\sigma _C$ due to angular clustering of the source galaxies, and errors in the 
limiting apparent
magnitude due to experimental errors and to uncertainties in the 
extinction/reddening correction. Let us start with the latter 
error in $m_{lim}$. If the galaxy counts scale as 
$N\sim 10^{s m_{lim}}$ with $s=0.15$, then
it is easy to show that the error in the number counts from an error in $
m_{lim}$ is smaller than the Poisson fluctuation only if $\Delta
m_{lim}<0.03 $. This is possible with current observational techniques; 
moreover, the amount of
extinction can in principle be estimated with multi-color data and by 
correlating with dust maps (though uncertainties in this procedure can 
cause systematic errors). Therefore, in the following we
will neglect the observational errors in $m_{lim}$ and focus on the other three
sources of noise, which are inevitably present independent of the data quality.
The total error in the number counts is therefore 
\begin{equation}
\sigma _{N}^2=\sigma _L^2+\sigma _P^2+\sigma _C^2 ~.
\end{equation}

The cosmic variance in the magnification field, the source of $\sigma_L$, 
is due to lensing 
by the random superposition of large-scale density fluctuations along the line of sight 
(this quantity is of considerable interest in its own right, and several 
lensing surveys are underway to detect it). The
variance of the convergence field $\kappa $ has been evaluated in Bernardeau 
{\it et al.} (1996). For a matter power spectrum $P(k)$ 
and an angular top-hat window function $W_{2D}$ of
size $\theta _0$, the variance of the
convergence field is given in the Einstein-de Sitter universe by  
\begin{equation}
\sigma _\kappa ^2=(2\pi )^{-1}\int_0^\infty dz\;f(z)\; w^2(z)\;D_{+}^2(z)\int_0^\infty k\;
P(k)\;W_{2D}^2[kr_p(z)\theta _0]\;dk ~, 
\end{equation}
where  $r_p(z)=r(0,z)(1+z)$, $f(z)=dr_p/dz$, 
 $D_{+}=(1+z)^{-1}$ is the growing mode of density fluctuations, and $w(z)
$ is an ``efficiency'' function which depends on the source distribution
(see Bernardeau {\it et al.} 1996): 
\begin{equation}
w(z)=\frac 32(1+z)r_p(z)\int_{z}^{\infty} p(z') \;
[1-r_p(z)/r_p(z^{\prime })]\;dz^{\prime }.
\end{equation}
For the density fluctuation power spectrum, we assume a cold dark matter (CDM) 
model with shape parameter $\Gamma =0.25$ and normalization $\sigma _8=0.6$ 
(White, Efstathiou \& Frenk
1993); we use the Bardeen et al. (1986) transfer function. Using Eqn.(20) 
and the relation between $\mu$ and $\kappa$, we have 
\begin{equation}
\sigma _L^2\approx (2-5 s )^2\sigma _\kappa ^2.
\end{equation}

Another source of statistical error is the variance due to the angular clustering of
the background sources. In terms of the present power spectrum $P(k)$, the variance of the
density contrast in circular patches of angular radius $\theta $ can be obtained by
writing the angular density contrast in an area $A=\pi \theta ^2$ as 
\begin{equation}
\delta (\theta )=A^{-1}\int_Ad\Omega \int dz \;p(z)\;\delta
(r,z)\;D_{+}(z) ~,
\end{equation}
where $\delta(r,z)$ is the density field perturbation.
This gives 
\begin{equation}
\sigma _C^2= \langle [\delta (\theta )]^2\rangle
    =(2\pi )^{-1}\int_0^\infty dz\; p^2(z)\; D_{+}^2(z)
\int_0^\infty k\;P(k)\;W_{2D}^2[kr_p(z)\theta _0]\;dk.
\end{equation}
Finally, the Poisson noise is readily estimated as $\sigma _P^2=AN_0$, where $A$
is the area of the field in square degrees.

\begin{figure}
\psfig{file=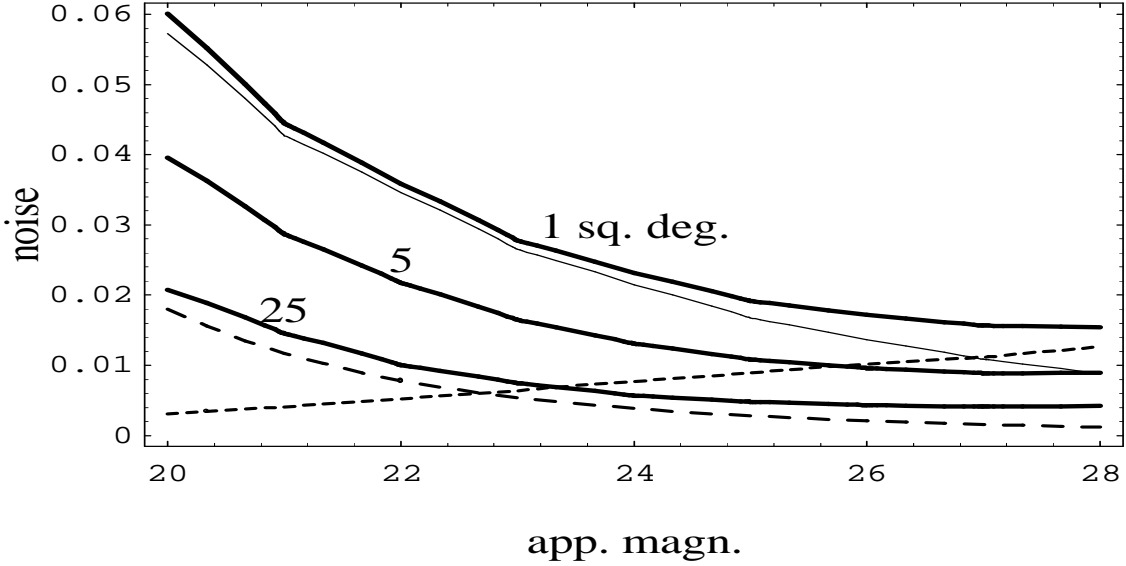,height=7.5cm,width=15cm}
\caption{The various sources of error in the number of galaxies observed to 
limiting $R$ magnitude over a given area behind a void: Poisson
noise $\sigma^2_P$ for 1 deg$^2$ field (long dashed curve); lensing variance 
$\sigma_L^2$ for 1 deg$^2$
(short dashed line); clustering variance $\sigma_C^2$ for 1 deg$^2$ (thin line); total
variance $\sigma_N^2$ for 1 deg$^2$ (top thick curve),  5 deg$^2$ (intermediate thick
curve), and 25 deg$^2$ (bottom thick curve).}
\end{figure}

Clearly, the Poisson and clustering contributions {\it decrease} with increasing 
depth of the sample, because the number of sources grows and the angular 
correlation function falls; by contrast, the lensing
variance $\sigma_L^2$ {\it increases } with sample depth, because the number of
lenses increases. Consequently, there is a depth for
which  $\sigma _N$ is a minimum. Of course, all the error contributions 
are smaller for a sample of larger angular size $\theta$. For a void at 
fixed redshift $z_L$, the lensing signal 
$\delta _b$ reaches a constant value at large depths, $z_S \gg z_L$, 
because the lensing effect is
maximal for sources at $r(0,z_S)\approx 2r(0,z_L)$. We therefore expect 
the lens signal-to-noise to peak at a depth for which the 
noise is minimal; we confirm this below for the 
values of $z_L$ we have studied.

To estimate the signal-to-noise we must 
specify the source galaxy distribution $p(z)$. We consider two different 
models for the galaxy redshift distribution in a deep magnitude-limited sample.
In the first, we follow Broadhurst, Taylor, \& Peacock (1995) 
in assuming an evolving Schechter (1976) luminosity function,  
\begin{equation}
\Phi (L,z)=\Phi ^{*}(z)\exp [-L/L_{*}]~,
\end{equation}
with $\Phi ^{*}(z)=0.02h^{-1}(1+z)^2(h^{-1}{\rm ~Mpc})^{-3}$. The value $
M^{*}=-21.5$ (for $h=1$) and a $K$-correction $K(z)=5\log _{10}(1+z)$ are
appropriate values in the $R$ band. The fraction of galaxies in the redshift 
interval ($z,z+dz$) observed in a sample of $N_0$ galaxies with limiting apparent
magnitude $m_{lim}$ (or limiting luminosity $L_{lim}(z_S)$), is given by 
\begin{equation}
p(z)dz=N_0^{-1}r^2_p(z)f(z)dz\int_{L_{lim}}^\infty \Phi (L,z)dL  \label{psi}
\end{equation}
where 
\begin{equation}
L_{lim}=L_{\star }10^{0.4\left[ M_{\star }-m_{lim}+25+5\log
[(1+z_S)^2r_p(z_S)]\right] }~.
\end{equation}
The second model provides a direct analytic parameterization of the redshift distribution  
based on a fit to galaxy redshift data 
 (Efstathiou et al. 1991; Brainerd et al. 1995), 
\begin{equation}
p(z)={3\over z_0^3 \Gamma(1+3/\beta)}z^2 e^{-(z/z_0)^{\beta}} ~, \label{pz}
\end{equation}
where $z_0$ and $\beta$ depend on the limiting magnitude. Empirically, 
$z_0=0.7$ and $\beta=2.6$ provides a good fit for  
$r_{lim}\sim 22$. For $r_{lim}\sim 22.5$, one can use $z_0=0.8$ and the same value of 
$\beta$. To obtain a typical estimate for 
deeper samples, we will use $z_0=1$ and $\beta=1.5$ 
(see, e.g., Schneider 1996); in this case, the median redshift of 
the sample is $\langle z\rangle \simeq 1.5$.

In Fig. 5  we show the sources of statistical error $\sigma
_P,\sigma _C,$ $\sigma _L$, and $\sigma _{N}$ as a function of $m_{lim}$ (limiting 
$R$ magnitude) 
for various sample areas, assuming the redshift distribution of Eq. (\ref{psi}). As 
anticipated, the clustering variance dominates
the other sources of error down to $m_{\lim }=26$, where the lensing variance becomes
comparable. In Fig. 6a we plot the signal-to-noise ratio $S/N\equiv \delta _b\,/\sigma
_{N}$ for void parameters $\tilde{d}=0.1$, $\tilde{b}=0.1$, and void radius $R=30$ \hm, 
as a function of void redshift $z_L$ and limiting magnitude $m_{lim}$, for a 
sample of area 25 deg$^2$. (As expected from Fig. 2, we 
find that the results do not change appreciably 
for $\tilde{d}<0.1$ and $\tilde{b}<0.1$.) It is evident that 
$S/N$ is well below unity at all limiting magnitudes and void redshifts. 
Since $S/N$ is
linear in void radius, we conclude that only a huge, empty
void of radius $R > 100$ \hm would be detectable via lensing magnification. 
The same conclusion can be drawn from Fig. 6b, in which we plot $S/N$ 
versus $z_L$ for the
same void parameters as above, but assuming the source redshift distribution of
Eq. (\ref{pz}) with $z_0=1$ and $\beta=1.5$.
The effect of several voids along
the line of sight could accumulate to give a larger flux magnification, if no
significant cluster is also intersected. However, as we have shown, for any
given $m_{lim}$, there is a maximal contribution for a specific value of $
z_L $, so that other voids far from this particular redshift are subdominant.

\begin{figure}
\hspace*{-0.1in}
\psfig{file=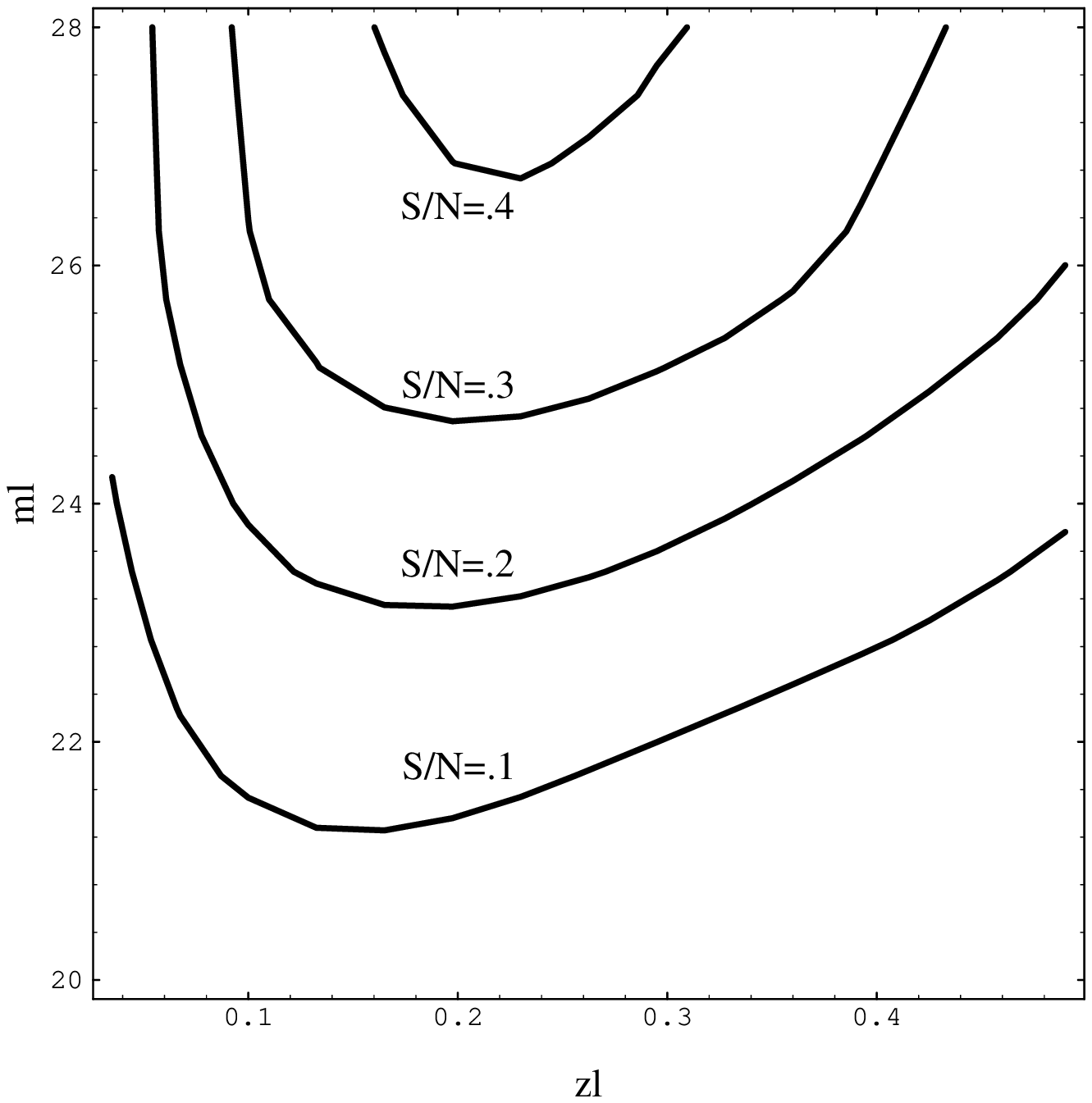,height=6.5cm,width=6.5cm}
\hspace*{0.2in}
\psfig{file=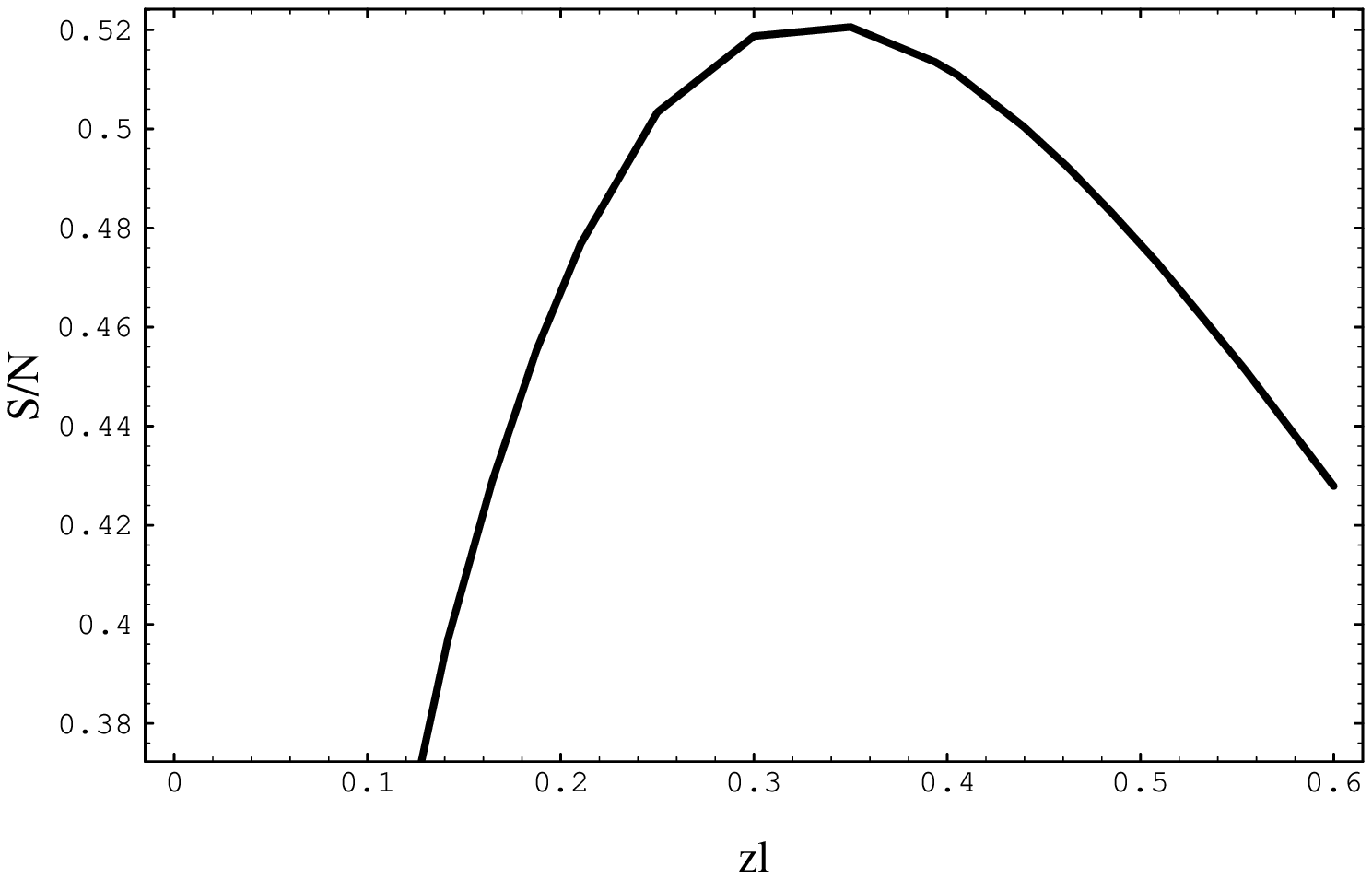,height=6.5cm,width=10cm}
\caption{The signal-to-noise ratio $S/N$ for $\tilde{d}=0.1$, $\tilde{b}=0.1$ and
void radius $R=30$ \hm as a function of $z_L$ and $m_l$, assuming $s =0.15$, for a 
sample of area of 25 deg$^2$ (left panel); the right panel shows $S/N$
versus $z_L$ for the same parameters, but assuming the redshift distribution of
Eq. (\ref{pz}) with $z_0=1$ and $\beta=1.5$.}
\end{figure}

\section{Aperture densitometry}

In this section we apply the technique of aperture densitometry to the detection 
of voids. In this method,  
proposed by Kaiser ($1995$) and generalized by Schneider ($1996$), an 
upper limit to the projected mass of a lens inside an aperture is derived by measuring the 
shear inside an annulus. For rays passing well inside the void boundary, 
the convergence is negative; as a result, this technique will yield  
a lower limit to the void mass deficit.

Following Kaiser (1995) and
Schneider (1996) we define the aperture mass, 
\begin{equation}
m(\theta_1, \theta_2) = \int d^2x \;\kappa(\vec{\theta}) \;w_k(\theta),
\end{equation}
and we use the Kaiser compensated filter function, 
\begin{eqnarray}
w_k (\theta) = \left\{ 
\begin{array}{lll}
\frac{1}{\pi \theta_1^2} & \;\;{\rm for}\;\;\;\;\;\;\;\;\;\;\; \theta < \theta_1, &  \\ 
\frac{-1}{\pi(\theta_2^2-\theta_1^2)} & \;\;{\rm for}\;\;\;\; \theta_1\leq \theta<\theta_2, &  \\ 
0 & \;\;{\rm for}\;\; \;\;\;\;\;\;\;\;\;\; \theta\geq \theta_2, & 
\end{array}
\right.\label{wk}
\end{eqnarray}
which satisfies the normalization condition $\int_0^{\theta_2} d\theta \; \theta \;
 w_k(\theta) =0$.

The aperture mass can be expressed in terms of the shear, 
\begin{equation}
m(\theta_1, \theta_2) = \int d^2x \;\frac{\gamma_t(\vec{\theta})}{\theta^2}q(\theta),
\label{m12}
\end{equation}
where 
\begin{equation}
\gamma_t(\vec{\theta})= - \gamma_1(\vec{\theta}) \cos 2\phi + \gamma_2(\vec{\theta}) \sin
2\phi ,
\end{equation}
is the tangential shear at the angular position 
$\vec{\theta}=(\theta \cos \phi, \theta \sin \phi)$
relative to the void center, and the function $q(\theta)$, defined by 
\begin{equation}
q(\theta)= 2\int_0^\theta d\theta'\; \theta' \;w_k(\theta') - \; \;\theta^2 w_k(\theta),
\end{equation}
is equal to $\theta_2^2/\pi(\theta_2^2-\theta_1^2)$ for $\theta_1 \leq \theta < 
\theta_2$ and zero
outside this region. Since $\langle \gamma _t \rangle = 1/(2 \pi) \int
\gamma _t d\phi$, using (\ref{m12}) we recover 
the Kaiser $\zeta$-statistic (Kaiser 1995), 
\begin{equation}
m(\theta_1, \theta_2) = \frac{2}{\left(1-\frac{\theta_1 ^2}{\theta_2 ^2}\right)} 
\int_{\theta_1}^{\theta_2} d\theta\, 
\frac{\langle \gamma _t \rangle}{\theta} = \overline{\kappa}(\theta<\theta_1) - 
\overline{\kappa}
(\theta_1<\theta<\theta_2).
\end{equation}
Here, $\overline{\kappa} = \int d^2\theta\;\kappa /\int d^2\theta$ is
the mean mass density within the aperture. We are considering axially 
symmetric voids, for which $\langle\gamma_t\rangle = \gamma_t = \gamma$.
Further, since $\gamma<0$, and away from the void boundary $\overline{\kappa}$ is also
negative, the shear inside the annulus can in principle be used to put a
lower bound on the mass deficit (as compared with the mean mass density of
the Universe) in the interior of a circle of radius $\theta_1$ around the center
of the void.

In our computations we assume the source galaxies are distributed in redshift
according to the normalized distribution function of Eqn.(\ref{pz}), 
and we take the values $z_0=1$ and $\beta=3/2$ as representative. 
The shear depends on the source redshift through the
ratio $D_{LS}/D_{OS}$, so we define the source distance average ratio, 
\begin{equation}
\left\langle \frac{D_{LS}}{D_S} \right\rangle= \int_{z_L}^\infty dz_S\; p(z_S) \;\frac{
D_{LS}}{D_{OS}}(z_L,z_S).
\end{equation}

A discretised estimator for the aperture mass (\ref{m12}) is given by 
\begin{equation}
m = \frac1n \sum_i \varepsilon_t (\vec{\theta}_i) \frac{q(\theta_i)}{\theta_i^2}~,\label{m}
\end{equation}
where $n$ is the source galaxy surface number density, and $\varepsilon_t$ is the 
tangential
ellipticity of a galaxy at position $\vec{\theta}_i$. Since the intrinsic
orientations of the source galaxies are random, in the absence of lensing 
$\langle m \rangle =0$. Squaring (\ref{m}) and taking the expectation
value, the dispersion of $m$ is (Schneider 1996) 
\begin{equation}
\sigma_d^2 = \frac{\sigma_{\varepsilon}^2}{2n^2}\sum_i \frac{q^2(\theta_i)}{\theta_i^4}
,\label{sigd}
\end{equation}
where $\sigma_{\varepsilon}=\sqrt{\langle |\varepsilon|^2\rangle}$ is the
dispersion of the intrinsic ellipticity distribution.
The expectation value of the aperture mass estimator is  
\begin{equation}
\langle m\rangle _d = \frac1n \sum_i \gamma_t (\vec{\theta}_i) \frac{q(\theta_i)}{\theta_i^2
}~.
\end{equation}

Performing an ensemble average of (\ref{m}) and (\ref{sigd}) over the probability
distribution for the galaxy positions we get (Schneider 1996) 
\begin{equation}
\langle m \rangle _c = 2 \pi \int_{\theta_1}^{\theta_2} d\theta \, \langle \gamma _t
\rangle\, \frac{q(\theta)}{\theta}
\end{equation}
and 
\begin{equation}
\sigma _c ^2(\theta_1, \theta_2) = \frac{\pi \sigma_\varepsilon ^2}{n}
\int_{\theta_1}^{\theta_2} d\theta \,\frac{q^2(\theta)}{\theta^3}.
\end{equation}
Using  $\bar{\rho}=\bar{\rho}_0\;(1+z_L)^3$ and
defining the scaled angle 
$\tilde{\theta}= \theta/\theta_R$, with $\theta_R \equiv R/D_{OL}$ the 
angular size of the void, we obtain the
ensemble-average signal-to noise ratio (for the Einstein-de Sitter model)
\begin{equation}
S_c = \frac{|\langle m \rangle _c|}{\sigma _c(\theta_1, \theta_2)} \simeq 42 \;\delta
\;(1+z_L)^3\;\left(\frac{0.2}{\sigma_{\varepsilon}}\right) \;\left(\frac{n}{30~
{\rm arcmin}^{-2}}\right)^{1/2} \;\left(\frac{R}{20~ h^{-1} {\rm Mpc}}\right)^2 \;
\left\langle \frac{D_{LS}}{D_{OS}} \right\rangle \;v(\tilde{d},\tilde{\theta}_1,
\tilde{\theta}_2),
\end{equation}
where the dimensionless function 
\begin{equation}
v(\tilde{d},\tilde{\theta}_1,\tilde{\theta}_2) \equiv \frac{\Sigma_c}{\Sigma_v}
\frac{\tilde{\theta}_1\;\tilde{\theta}_2}{\sqrt{\tilde{\theta}_2^2 - \tilde{\theta}_1^2}}
\int_{\tilde{\theta}_1}^{\tilde{\theta}_2} d{\tilde{\theta}}\; \frac{|\gamma_t
(\tilde{\theta},\tilde{d})|}{\tilde{\theta}}
\end{equation}
is typically of order unity or less.

For example, for an empty void ($\delta = 1$) of radius $R=30$ \hm and 
fractional shell thickness ${\tilde d}=0.1$ at 
redshift $z_L=0.1$, the signal-to-noise is $S_c \simeq 5$ for an 
ellipticity variance $\sigma_c = 0.2$, aperture thickness 
$\Delta \theta = \theta_2 - \theta_1 = 20$ arcmin (corresponding to 
$\Delta \tilde{\theta} = \tilde{\theta_2} - \tilde{\theta_1}=0.1$) and 
outer radius $\tilde{\theta_2} = 0.7$. Thus, 
taking into account only
the noise due to the dispersion of the intrinsic ellipticity distribution, we 
would conclude that voids can be detected with the aperture
densitometry technique. However, voids do not exist in an otherwise homogeneous
Universe: the mass is clustered on all scales, and the associated density 
perturbations cause random distortions of background galaxy images. The resulting 
cosmic variance in the aperture mass turns out to dominate over the intrinsic ellipticity
dispersion and makes void lensing extremely difficult to detect. To see this, we note 
that the signal to noise
ratio due to the intrinsic ellipticity distribution increases
with the area of the sample: to achieve large signal to noise, 
it is necessary to increase the area of the annulus. However, as we show 
below, a larger 
sample area enhances the dispersion due to rms density perturbations (cosmic 
variance), with the result that only very large voids yield an 
observable lensing effect.

To estimate the cosmic variance noise we follow the method described in
Schneider {\it et al.} (1998). Using the 
filter function of Eq.(\ref{wk}), we have 

\begin{equation}
\langle M_{ap}^2(\theta_1,\theta_2)\rangle = 2 \pi \int_0^\infty ds\;s\, P_k(s) \,[I(s
\theta_1,s \theta_2)]^2,\label{map}
\end{equation}
where 
\begin{equation}
I(s \theta_1,s \theta_2)= \frac{\theta_2}{s \pi \theta_1 (\theta_2^2-\theta_1^2)}
\left(\frac{\theta_2}{\theta_1} J_1(s\theta_1) - J_1(s\theta_2)\right) ~,
\end{equation}
with $J_n(x)$ denoting the Bessel function of the first kind. In Eq.(\ref{map}), the
quantity $P_k(s)$ is given by (Kaiser 1995, Schneider, {\it et al.} 1998), 
\begin{equation}
P_k(s)= \frac94 \left({c\over H_0}\right)^{-3} 
\int_0^\infty dz\;\left\langle\frac{D_{LS}}{D_{OS}}
\right\rangle(z)\;(1+z)^{-\frac{3}{2}}\;P_0(s/r(0,z)),
\end{equation}
where $P_0(k)$ is the mass power spectrum. In our computation we use the
non-linear CDM power spectrum (Hamilton, {\it et al.} 1991, Peacock \& Dodds
1994,1996) with the Bardeen {\it et al.} (1986) transfer function, with shape
parameter $\Gamma=0.25$ and normalized such that $\sigma_8=0.6$. The ratio
between the cosmic variance and intrinsic ellipticity
distribution contribution to the noise is (Schneider {\it et al.} 1998): 
\begin{equation}
\rho= 1500 \pi \;\left(\frac{0.2}{\sigma_{\varepsilon}}\right)^2 \;
\left(\frac{n}{30~
{\rm arcmin}^{-2}}\right) \; \langle M_{ap}^2(\theta_1,\theta_2)\rangle \;
\left(\frac{\theta_1}{{\rm arcmin}}\right)^{2}
\left(1-\frac{{\theta_1}^2}{{\theta_2}^2}\right) ~.
\end{equation} 
The total aperture densitometry signal to noise ratio is then 
\begin{equation}
\frac{S}{N} = \frac{S_c}{\sqrt{1+\rho^2}}~.
\end{equation}
\begin{figure}
\hspace*{0.1in} \vspace*{0.5cm} 
\psfig{file=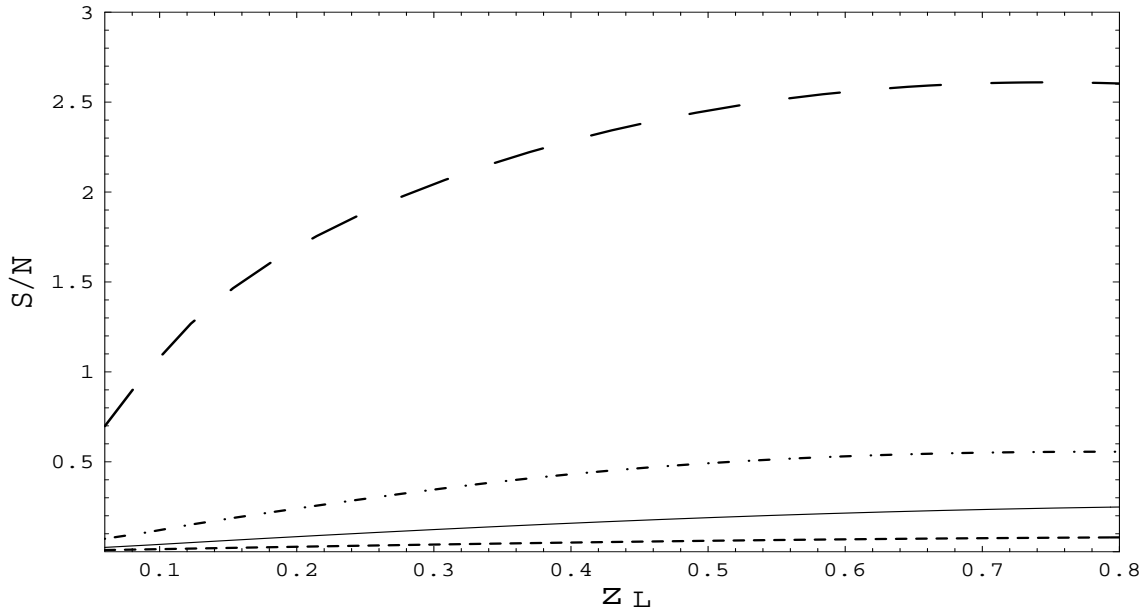,height=8cm,width=15cm}
\vspace*{0.1in}
\caption{Aperture densitometry signal to noise ratio 
 as a function of void redshift $z_L$, for  
an empty ($\delta=1$) void of radius $R = 30 h^{-1}$ Mpc and 
shell thickness $\tilde{d}=0.1$, assuming a density of background galaxies $n=30$
arcmin $^{-2}$ with ellipticity dispersion $\sigma_{\varepsilon}=0.2$, evaluated 
at an outer angular radius $\tilde{\theta}_2 = 0.7$, and assuming a $\Gamma=0.2$ 
non-linear CDM power spectrum for the mass fluctuations. Results are shown for 
aperture width $\Delta \theta \equiv \theta_2 - \theta_1 \simeq 20$ (dot dashed), $2$ (solid), and 
$0.2$ arcmin (dashed) 
(corresponding to $\Delta \tilde{\theta}=0.1,\;0.01$, and $10^{-3}$). The long 
dashed curve shows results for the same model parameters but for a void of  
radius $R = 100$ \hm and  aperture $\Delta \tilde{\theta}=10^{-4}$.}
\end{figure}

In Fig. $7$ we plot the aperture densitometry signal to noise ratio as a function of void
redshift $z_L$ for a void of radius $R=30$ \hm. 
We also show results for  $R = 100$ \hm, and  
aperture $\Delta \tilde{\theta}=10^{-4}$. This shows that very large void 
radii are required to achieve a signal to noise ratio larger than unity.
We conclude that the void lensing signal for individual objects 
is quite difficult to detect via
 aperture densitometry, due to the large contribution of cosmic variance. 

\section{Summary and discussion}

We have studied two methods for detecting weak lensing by large-scale 
voids in the matter distribution, color-dependent density magnification
and aperture densitometry. We have found that only voids larger
than $R \sim$ 100 \hm have a chance of being observed via lensing.
In both cases, the limiting factor is clustering: for 
density magnification, the problem is angular clustering 
of the background sources; for aperture densitometry, 
the limit is set by lensing due to other clustered mass along 
the line of sight.

We note that the model studied here is highly idealized. In particular, 
we have assumed that a single void lens dominates the distortion 
and amplification along the line of sight, while typical light rays from 
high-redshift sources will in fact traverse a number of voids. On the 
other hand, if voids are ubiquitous, their mean cumulative 
effects should be described by the large-scale power spectrum, 
which we have taken into account via the cosmic variance noise.
Our study therefore should be considered as applying 
to the detectability of individual, 
relatively rare, large, nearly empty voids embedded in a
matter distribution otherwise 
described by a CDM-like power spectrum. The existence
of such large voids in the galaxy distribution 
is a question that should be addressed by the next generation
of large redshift surveys, including the Sloan Digital Sky Survey 
and the Two Degree Field Survey.

\section*{Acknowledgments}

We thank Peter Schneider for helpful suggestions and discussions. I. W. was 
supported in part by the Brazilian agencies CNPq and FAPERJ. This
work was supported by the DOE and NASA Grant NAG5-7092 at Fermilab.

\section*{References}

\parskip 0pt\noindent\hangindent 20pt Amendola L. \& Occhionero F. , 1993,
Ap. J., {\bf 413}, 39.

\parskip 0pt\noindent\hangindent 20pt Amendola L. {\it et al.}, 1996, Phys.
Rev. D{\bf 54}, 7150.

\parskip 0pt\noindent\hangindent 20pt Bardeen J. M., Bond J. R., Kaiser N.,
Szalay A. S., 1986, Ap. J., {\bf 304}, 15.

\parskip 0pt\noindent\hangindent 20pt Bernardeau F. , Van Waerbeke L.,
Mellier Y., 1997,  A\& A, {\bf 322}, 1.

\parskip 0pt\noindent\hangindent 20pt Broadhurst T. J., Taylor A. N. \&
Peacock J. A., Ap. J., 1995, {\bf 438}, 49. 

\parskip 0pt\noindent\hangindent 20pt Broadhurst T. J., 1996,
``Gravitational `Convergence' and Galaxy Cluster Masses''- astro-ph/9511150.

\parskip 0pt\noindent\hangindent 20pt da Costa L. N. {\it et al.} 1996, Ap.
J., {\bf 468}, L5.

\parskip 0pt\noindent\hangindent 20pt de Lapparent V., Geller M., \& Huchra
J., 1989, Ap. J., {\bf 343}, 1.

\parskip 0pt\noindent\hangindent 20pt Dyer, C. C. \& Roeder R. C. 1972, Ap. J., 
{\bf 174 }, L115.

\parskip 0pt\noindent\hangindent 20pt El-Ad H., Piran T. and da Costa L. N.,
1996, Ap. J., {\bf 462}, L13.

\parskip 0pt\noindent\hangindent 20pt El-Ad H., Piran T. and da Costa L. N.,
1997,MNRAS, {\bf 287}, 790.

\parskip 0pt\noindent\hangindent 20pt Hamilton A. J. S., Kumar P., Lu E.,
Matthews A., 1991, Ap. J., {\bf 374}, L1.

\parskip 0pt\noindent\hangindent 20pt Hammer F. \& Nottale L., 1986, A\& A, 
{\bf 167}, 1.

\parskip 0pt\noindent\hangindent 20pt Holz D. \& Wald R. 1998, 
Phys. Rev. D{\bf 58}, 063501.

\parskip 0pt\noindent\hangindent 20pt Kaiser N., 1995, Ap. J., {\bf 439}, L1. 

\parskip 0pt\noindent\hangindent 20pt La D., 1991, Phys. Lett. B {\bf 265},
232.

\parskip 0pt\noindent\hangindent 20pt Moreno J. \& Portilla M., 1990, Ap.
J., {\bf 352}, 399.

\parskip 0pt\noindent\hangindent 20pt Ostriker J. P. \& Cowie L.N., 1981,
Ap. J., {\bf 243}, L127.

\parskip 0pt\noindent\hangindent 20pt Peacock J. A., Dodds S. J., 1994,
MNRAS, {\bf 267}, 1020.

\parskip 0pt\noindent\hangindent 20pt Peacock J. A., Dodds S. J., 1996,
MNRAS, {\bf 280}, L19.

\parskip 0pt\noindent\hangindent 20pt Schechter P., 1976,
Ap. J., {\bf 203}, 297.

\parskip 0pt\noindent\hangindent 20pt Schneider P., 1996, MNRAS, {\bf 283},
837. 

\parskip 0pt\noindent\hangindent 20pt Schneider P., Ehlers J. \& Falco E.
E., 1992,{\it Gravitational Lenses}, Springer: New York.

\parskip 0pt\noindent\hangindent 20pt Schneider P., van Waerbeke L., Jain B.
and Kruse G., 1998, MNRAS, {\bf 296}, 873. 

\parskip 0pt\noindent\hangindent 20pt Van Kampen E., 1998, astro-ph/9807305

\parskip 0pt\noindent\hangindent 20pt Yoshioka S. and Ikeuchi S., 1989, Ap.
J., {\bf 341}, 16.

\parskip 0pt\noindent\hangindent 20pt White S. D. M., Efstathiou G. and
Frenk C. S., 1993, MNRAS. {\bf 262},1023.

\end{document}